\documentclass[runningheads]{llncs}
\usepackage{amssymb,amsmath}
\usepackage{pstricks}
\usepackage{pst-node}
\usepackage{epsfig}
\usepackage{latexsym}
\begin{document}
\pagestyle{headings}

\newcommand{\tit}[1]{{\textit{#1\/}}}
\makeatletter
\DeclareRobustCommand\xspace{\futurelet\@let@token\@xspace}
\def\@xspace{%
  \ifx\@let@token\bgroup\else
  \ifx\@let@token\egroup\else
  \ifx\@let@token\/\else
  \ifx\@let@token\ \else
  \ifx\@let@token~\else
  \ifx\@let@token.\else
  \ifx\@let@token!\else
  \ifx\@let@token,\else
  \ifx\@let@token:\else
  \ifx\@let@token;\else
  \ifx\@let@token?\else
  \ifx\@let@token/\else
  \ifx\@let@token'\else
  \ifx\@let@token)\else
  \ifx\@let@token-\else
  \ifx\@let@token\@xobeysp\else
  \ifx\@let@token\space\else
  \ifx\@let@token\@sptoken\else
   \space
   \fi\fi\fi\fi\fi\fi\fi\fi\fi\fi\fi\fi\fi\fi\fi\fi\fi\fi}
\makeatother

\newcommand{\mt}[1]{\mathtt{#1}}
\newcommand{\seq}[2][n]{\ensuremath{#2_{1},\dots,#2_{#1}}\xspace}

\newcounter{clause}
\def\theclause{$c$\arabic{clause}}

\newenvironment{clause}{\begin{tabbing}
xxx\=xxx\=xxx\=\+\kill}%
{\end{tabbing}}

\newenvironment{Clause}{\refstepcounter{clause}%
\begin{tabbing}
cxxxx\=xxx\=xxx\=\kill
\theclause\>\+}%
{\end{tabbing}}

\let\oldP\P
\newcommand{\M}{\ensuremath{\mathcal{M}}\xspace}
\newcommand{\D}{\ensuremath{\mathcal{D}}\xspace}
\newcommand{\Prog}{\ensuremath{\mathcal{P}}\xspace}
\let\P\Prog

\newcommand{\finish}[1]{}
\newcommand{\ignore}[1]{}
\renewcommand{\labelitemi}{\mbox{$\bullet$}}
\newcommand{\ar}{\hookrightarrow}

\newcommand{\Vars}{{\it Vars}}
\newcommand{\Valn}{{\it Valn}}
\newcommand{\dft}{{\it dft}}
\newcommand{\MT}{\mbox{$\M_\T$}}

\renewcommand{\H}{{\cal H}}
\newcommand{\I}{{\cal I}}
\newcommand{\T}{{\cal T}}
\newcommand{\U}{{\cal U}}
\newcommand{\W}{{\cal W}}
\renewcommand{\d}{\tilde{d}}
\newcommand{\x}{\tilde{x}}

\newcommand{\lsem}{{[\![}}
\newcommand{\rsem}{{]\!]}}

\ignore{
\newcommand{\qed}{
    \setlength{\unitlength}{6pt}
    \begin{picture}(1,1)
    \thicklines
    \multiput(0,0)(0.0,0.01){100}{\line(1,0){1}}
    \end{picture}
}

\newtheorem{theorem}{Theorem}[section]
\newtheorem{lemma}{Lemma}[section]
\newtheorem{definition}{Definition}[section]
\newtheorem{defn}{Definition}[section]
\newtheorem{corollary}{Corollary}[section]
\newtheorem{proposition}{Proposition}[section]
\newtheorem{propn}{Proposition}[section]
\newtheorem{auxexample}{Example}[section]
\newenvironment{example}{\begin{auxexample}\em }{\ $\Box$\end{auxexample}}
\newenvironment{proof}{\begin{quotation} \noindent \em
	   {\bf Proof:\ }}{\(\Box\) \end{quotation}}
}

\newtheorem{propn}{Proposition}[section]

\title{A Model-Theoretic Semantics\\
for Defeasible Logic
\thanks{Originally published in proc. PCL 2002, a FLoC workshop;
eds. Hendrik Decker, Dina Goldin, J{\o}rgen Villadsen, Toshiharu Waragai
({\tt http://floc02.diku.dk/PCL/}).}
}

\author{Michael J. Maher}
\institute{
Department of Computer Science \\
Loyola University Chicago \\
{\tt mjm@cs.luc.edu} \\
\ \\
CIT, Griffith University \\
Nathan, QLD 4111, Australia \\
}
\date{}

\maketitle

\newcommand{\pd}[1]{+\partial #1}
\newcommand{\md}[1]{-\partial #1}
\newcommand{\PD}[1]{+\Delta #1}
\newcommand{\MD}[1]{-\Delta #1}
\newcommand{\zd}[1]{\overline{\md{#1}}}
\newcommand{\ZD}[1]{\overline{\MD{#1}}}
\newcommand{\id}[1]{\infty\partial #1}

\newcommand{\nul}{\mbox{{\em $\langle\!\!$ null $\!\rangle$}}}
\newcommand{\supr}{\mbox{$\!>\!$}}
\newcommand{\p}{{\em ~~Poss~~}}

\newcommand{\non}{\!\sim\!\!}

\newcommand{\Conc}{{\bf Conc}}
\renewcommand{\P}{{\cal P}}

\begin{abstract}
Defeasible logic is an efficient logic for defeasible reasoning.
It is defined through a proof theory and, until now, has had no model theory.
In this paper 
a model-theoretic semantics is given for defeasible logic.
The logic is sound and complete with respect to the semantics.
We also briefly outline how this approach extends to a wide range
of defeasible logics.
\end{abstract}

\finish{RESET finish}

\section{Introduction}

Defeasible logic is a logic designed for efficient defeasible reasoning.
The logic was designed by Nute \cite{Nute88,Nute}
with the intention that it be efficiently implementable.
This intention has been realised in systems that can process
hundreds of thousands of defeasible rules quickly \cite{ICTAI}.
Over the years, Nute and others have proposed many variants of defeasible logic
\cite{Nute,ABGM}.
In this paper we will address a particular defeasible logic,
which we denote by \tit{DL} \cite{Billington,TOCL}.
However, this work is easily modified to address other defeasible logics.

Defeasible logics are, in general, paraconsistent.
The nature of defeasible reasoning,
where one chain of reasoning may defeat another,
predisposes the logics to avoid inconsistent inferences,
and provides for a natural treatment of inconsistencies, when they occur.
In the case of \tit{DL}, which supports only sceptical reasoning,
inconsistent inferences are almost completely avoided.
They only occur as a result of inconsistencies in the
definite knowledge expressed by a theory.
When such inconsistencies occur, 
the inconsistent literals may be used individually -- or even together --
in further inferences, but no form of {\em ex falso quodlibet}
reasoning is possible.

\tit{DL}, and similar logics, have been proposed as the
appropriate language for executable regulations \cite{hicss},
negotiations \cite{GDHO},
contracts \cite{Grosof1}, and business rules \cite{Grosof2}.
The logics are considered to have satisfactory expressiveness
and the efficiency of the implementations supports real-time response
in applications such as electronic commerce \cite{Grosof2,collecter}.
Indeed, propositional \tit{DL} has been shown to have linear time complexity
\cite{linear}.

On the other hand, neither \tit{DL} nor any other variant of modern defeasible logic
has a model theory.
\tit{DL} is defined purely in proof-theoretic terms \cite{Nute,Billington}.
Furthermore, a model theory based on the idea of extensions
(such as is used for default logic) is
likely to be inappropriate for defeasible logic,
since the kind of scepticism that is developed from intersection of extensions
in default logic is different from the kind of ``direct'' scepticism \cite{Horty}
that occurs in defeasible logic \cite{Schlechta}.

In early work on semantics for defeasible logics, Nute \cite{Nute88}
defined a model theory for LDR, a substantially simpler precursor of \tit{DL},
in terms of a minimal belief state for each theory.
LDR defines defeat only in terms of definite provability;
this limitation is the main reason why the approach is successful \cite{Nute88}.
Recently, this approach has been extended \cite{Donnelly,ND}
to a defeasible logic that is closer to \tit{DL},
and more general in some respects.
However,
the semantics is based on the idea of intersection of extensions
and -- perhaps consequently --
the logic is sound but not complete for this semantics.

Although there is not a model theory for defeasible logic,
there has been work on providing a semantics for \tit{DL}
in other styles.
In \cite{MG99} we showed that \tit{DL} can be defined
in terms of a metaprogram, defined to reflect the inference rules
of the logic, and a semantics for the language of the metaprogram.
This approach was successful in establishing a relationship
between \tit{DL} and Kunen's semantics of negation-as-failure.
Although it did not directly address model-theoretic reasoning,
we will use this semantics as a key intermediate step
in verifying the correctness of our model-theoretic semantics.

We have described \tit{DL} and its variants in
argumentation-theoretic terms \cite{GM00,pricai}.
Such a characterization is useful for the applications
of the logic that we have in mind,
but the resulting semantics is again a meta-level treatment of the proof
theory:
proof trees are grouped together as arguments,
and conflicting arguments are resolved by notions of argument defeat
that reflect defeat in defeasible logic.
Thus this work also fails to address model theory.

Finally, a denotational-style semantics \cite{Scott} 
has been given to \tit{DL} \cite{densem}.
The semantics is compositional,
and fully abstract in all but one syntactic class.
Although this semantics provides a useful analysis of \tit{DL},
it does not provide a model theory.

In this paper, we define a model-theoretic semantics for \tit{DL}.
This semantics follows Nute's semantics for LDR in that
models represent a state of mind or ``belief state''
in which definite knowledge (that which is ``known'') is distinguished from
defeasible knowledge (that which is ``believed'').
A major difference from \cite{Nute88}
is that it adopts partial models as the basis from which to work.
The approach can be adapted easily to a wide range of defeasible logics.

The structure of the paper is as follows.
In the next section we introduce the constructs of defeasible logic,
and the proof theory of \tit{DL}.
We then introduce the model-theoretic semantics
and prove the soundness and completeness of the proof system
with respect to this semantics.
Finally, we discuss the extension of this work to other defeasible logics,
including logics which admit pre-defined relations --
constraints in the sense of constraint logic programming \cite{JM}.

\section{The Defeasible Logic \tit{DL}}  \label{sec:DL}

We begin by presenting the basic ingredients of defeasible logic.
A defeasible theory
contains five different kinds of knowledge: facts,
strict rules, defeasible rules, defeaters, and a
superiority relation.

{\em Facts} are indisputable statements,
for example, ``Tweety is an emu''.
This might be expressed as $\tit{emu}(\tit{tweety})$.

{\em Strict rules} are rules in the traditional sense:
whenever  the premises are indisputable (e.g.\
facts) then so is the conclusion. An example of a strict rule is ``Emus
are birds''. Written formally:
$$\tit{emu}(X) \rightarrow \tit{bird}(X).$$
{\em Defeasible rules} are rules that can be defeated by contrary
evidence. An example of such a rule is ``Birds typically
fly''; written formally: $$\tit{bird}(X) \Rightarrow \tit{flies}(X)$$
The idea is that if we know that something is a bird, then we may
conclude that it flies, {\em unless there is other evidence suggesting
that it may not fly}.

{\em Defeaters} are rules that cannot be used
to draw any conclusions. Their only use is to prevent some
conclusions. In other words, they are used to defeat some
defeasible rules by producing evidence to the contrary. An
example is ``If an animal is heavy then it might not be able to fly''. Formally:
$$\tit{heavy}(X) \leadsto \neg \tit{flies}(X)$$
The main point is that the information that an animal is heavy is 
not sufficient evidence to
conclude that it doesn't fly. It is only evidence that the
animal {\em may} not be able to fly.
In other words, we don't wish to
conclude $\neg \tit{flies}(\tit{tweety})$ if $\tit{heavy}(\tit{tweety})$, we simply want to prevent
a conclusion $\tit{flies}(\tit{tweety})$.

The {\em superiority relation} among rules is used to define
priorities among rules,
that is,
where one rule may override the conclusion of another rule.
For example,
given the defeasible rules

\[
\begin{array}{lrl}
r: & \tit{bird}(X) & \Rightarrow \tit{flies}(X) \\

r':  & \tit{brokenWing}(X) & \Rightarrow \neg \tit{flies}(X) \\

\end{array}
\]
\noindent
which contradict one another,  no conclusive decision can be
made about whether a bird with a broken wing can fly. But if we 
introduce a superiority relation $>$ with $r'>r$,
then we can indeed conclude that the bird cannot fly.
We assume that $>$ is acyclic.
It turns out that we only need to define the superiority relation
over rules with contradictory conclusions.

We now turn to a formal description of the language of defeasible logic
and the inference rules of the defeasible logic \tit{DL}.

We assume a signature $\Sigma$
containing predefined function and predicate symbols,
including $=$,
with their arities.
Initially, we will assume that it contains only constants and $=$.
Later we will consider the full generality of $\Sigma$.
We also assume a signature $\Pi$ containing the set of predicate symbols,
with their arities,
that are defined by a defeasible theory.

An {\em atom} has the form $p(t_1, \ldots, t_n)$
where $p \in \Pi$ has arity $n$ and $t_1, \ldots, t_n$ are terms.
A {\em literal} has the form $a$ or $\neg a$, where $a$ is an atom.

A {\em rule} $r$ consists of an optional label,
an {\em antecedent} (or {\em body}) $A(r)$
which is a finite set of literals, an arrow, 
and its {\em head}, which is a literal.
Given a set $R$ of rules, we denote
the set of all strict rules in $R$ by $R_s$,
the set of strict and defeasible rules in $R$ by $R_{\tit{sd}}$,
the set of defeasible rules in $R$ by $R_d$, and
the set of defeaters in $R$ by $R_{\dft}$.
$R[q]$ denotes the set of rules in $R$ with head $q$.
If $q$ is a literal, $\non q$ denotes
the complementary literal (if $q$ is a positive literal $p$ then $\non q$ is
$\neg p$; and if $q$ is $\neg p$, then $\non q$ is $p$).

A {\em defeasible theory} $\T$ is a triple $(F,R,>)$ where $F$ is a finite
set of literals (called {\em facts}), $R$ a finite set of rules, and
$>$ a superiority relation on the labels of $R$.

\begin{example} \label{bird}
We will use the following defeasible theory to demonstrate some elements
of defeasible logic.
We assume there are only the constants $\tit{ethel}$ and $\tit{tweety}$
in the language.
Let
$T_\tit{bird} = (F_\tit{bird}, R_\tit{bird}, >_\tit{bird})$
where:
$F_\tit{bird}$ is the set of facts
\[
\begin{array}{lrrl}
& & & \tit{emu}(\tit{ethel}). \\
& & & \tit{bird}(\tit{tweety}). \\
\end{array}
\]
$R_\tit{bird}$ is represented by the set of rule schemas
\[
\begin{array}{lrrl}
r_1: & \tit{emu}(X) & \rightarrow & \tit{bird}(X). \\
r_2: & \tit{bird}(X) & \Rightarrow & \tit{flies}(X). \\
r_3: & \tit{heavy}(X) & \leadsto & \neg \tit{flies}(X). \\
r_4: & \tit{brokenWing}(X) & \Rightarrow & \neg \tit{flies}(X). \\
r_5: & & \Rightarrow & \tit{heavy}(\tit{ethel}). \\
\end{array}
\]
and the superiority relation $>_\tit{bird}$
contains only $r_4 >_\tit{bird} r_2$.

The five rule schemas
give rise to
nine propositional rules
by instantiating each variable to $\tit{ethel}$ and $\tit{tweety}$ respectively.
\ignore{
}
Those rules, which make up $R_\tit{bird}$ are
\[
\begin{array}{lrrl}
r_{1,e}: & \tit{emu}(\tit{ethel}) & \rightarrow & \tit{bird}(\tit{ethel}). \\
r_{1,t}: & \tit{emu}(\tit{tweety}) & \rightarrow & \tit{bird}(\tit{tweety}). \\
r_{2,e}: & \tit{bird}(\tit{ethel}) & \Rightarrow & \tit{flies}(\tit{ethel}). \\
r_{2,t}: & \tit{bird}(\tit{tweety}) & \Rightarrow & \tit{flies}(\tit{tweety}). \\
r_{3,e}: & \tit{heavy}(\tit{ethel}) & \leadsto & \neg \tit{flies}(\tit{ethel}). \\
r_{3,t}: & \tit{heavy}(\tit{tweety}) & \leadsto & \neg \tit{flies}(\tit{tweety}). \\
r_{4,e}: & \tit{brokenWing}(\tit{ethel}) & \Rightarrow & \neg \tit{flies}(\tit{ethel}). \\
r_{4,t}: & \tit{brokenWing}(\tit{tweety}) & \Rightarrow & \neg \tit{flies}(\tit{tweety}). \\
r_{5}: & & \Rightarrow & \tit{heavy}(\tit{ethel}). \\
\end{array}
\]
The rules have been re-labelled purely to simplify later reference\footnote{
Without the re-labelling we would have different rules with the same label.
This is not a problem, formally, but might be confusing.
}.
As a result, the superiority relation becomes
$\{
r_{4,e}  >_\tit{bird} r_{2,e},
r_{4,e}  >_\tit{bird} r_{2,t},
r_{4,t}  >_\tit{bird} r_{2,e},
r_{4,t}  >_\tit{bird} r_{2,t}
\}
$.
As noted in earlier, the two statements
$r_{4,e}  >_\tit{bird} r_{2,t}$
and
$r_{4,t}  >_\tit{bird} r_{2,e}$
have no effect,
since they do not involve rules with conflicting heads.

In this defeasible theory,
$R_s = \{ r_{1,e}, r_{1,t} \}$
\,and\, $R_d[\neg \tit{flies}(\tit{tweety})] = \{ r_{3,t}, r_{4,t} \}$.
\end{example}

A {\em conclusion} of $\T$ is a tagged literal and
can have one of the following four forms:

\begin{itemize}

\item[~] $+\Delta q$, which is intended to mean that the literal $q$
is definitely provable in $\T$
(i.e., using only facts and strict rules).

\item[~] $-\Delta q$, which is intended to mean that we have proved 
that $q$ is not definitely provable in $\T$.

\item[~] $+\partial q$, which is intended to mean that $q$ is defeasibly
provable in $\T$.

\item[~] $-\partial q$ which is intended to mean that we have proved that
$q$ is not defeasibly provable in $\T$.

\end{itemize}

Thus, conclusions are meta-theoretical statements about provability.
They do not appear in a defeasible theory.
Notice the distinction between $-$, which is used only to express
failure-to-prove, and $\neg$, which expresses classical negation.
For example,
$\MD{\neg \tit{flies}(\tit{tweety})}$ means
that it has been proved that the negated proposition $\neg \tit{flies}(\tit{tweety})$
cannot be proved definitely in the defeasible theory.

Provability is based on the concept of a {\em
derivation} (or proof)  in $\T=(F,R,>)$. A derivation is a finite sequence 
$P=(P(1),\ldots P(n))$ of
conclusions constructed by inference rules.

The preceding descriptions and definitions are common to many 
variants of defeasible logic, although in some variants the limitation to
four tags and four forms of conclusion has been discarded \cite{ABGM}.
The inference rules we present below are specific to,
and characterize,
the defeasible logic \tit{DL}.
We follow the presentation of Billington \cite{Billington}.

There are four inference rules for \tit{DL}
(corresponding to the four kinds of conclusion)
that specify how a derivation can be extended.
The formulation of these inference rules assumes
a propositional defeasible theory.
($P(1..i)$ denotes the initial part of the sequence
$P$ of length $i$):

\begin{tabbing}
90123456\=7890\=1234\=5678\=9012\=3456\=\kill

\>$+\Delta$: We may append $P(i+1) = +\Delta q$ if either \\
\>\>\> $q\in F$ or \\
\>\>\> $\exists r\in R_s[q] \ \forall a\in A(r): +\Delta a \in
P(1..i)$ 
\end{tabbing}

This means, to prove $+\Delta q$ we need to establish a proof for $q$
using facts and strict rules only. This is a deduction in the
classical sense.  To prove $-\Delta q$ it is required to show
that every attempt to prove $+\Delta q$ fails in a finite time.
Thus the inference rule for $-\Delta$ is the constructive complement
of the inference rule for $+\Delta$ \cite{MG99}.

\begin{tabbing}
90123456\=7890\=1234\=5678\=9012\=3456\=\kill

\> $-\Delta$: We may append $P(i+1)=-\Delta q$ if \\
\>\>\> $q\not\in F$ and \\
\>\>\> $\forall r\in R_{s}[q] \ \exists a\in A(r): -\Delta a \in P(1..i)$
\end{tabbing}

From $T_\tit{bird}$ in Example \ref{bird}
we can infer $\PD{\tit{emu}(\tit{ethel})}$ (and $\PD{\tit{bird}(\tit{tweety})}$) immediately,
in a proof of length 1.
Using $r_{1,e}$ and the second clause of the inference rule,
we can infer $\PD{\tit{bird}(\tit{ethel})}$ in a proof of length 2.
We can infer 
$\MD{\tit{heavy}(\tit{tweety})}$ and $\MD{\neg \tit{flies}(\tit{tweety})}$
immediately (among many others), since in these cases $R_{s}[q]$ is empty.

The inference rule for defeasible conclusions is complicated
by the defeasible nature of \tit{DL}:
opposing chains of reasoning must be taken into account.

\begin{tabbing}
90123456\=7890\=1234\=5678\=9012\=3456\=\kill
\>$+\partial$: \> We may append $P(i+1)=+\partial q$ if either \\
\>\>(1) $+\Delta q \in P(1..i)$ or \\
\>\>(2) \> (2.1) $\exists r\in R_{\tit{sd}}[q] \forall a \in
A(r): +\partial a\in P(1..i)$ and \\
\>\>\>(2.2) $-\Delta\non q \in P(1..i)$ and \\
\>\>\>(2.3) $\forall s \in R[\non q]$ either \\
\>\>\>\>(2.3.1) $\exists a\in A(s): -\partial a\in P(1..i)$ or \\
\>\>\>\>(2.3.2) $\exists t\in R_{\tit{sd}}[q]$ such that \\
\>\>\>\>\> $\forall a\in A(t): +\partial a\in P(1..i)$ and $t>s$ 
\end{tabbing}

Let us work through this inference rule.
To show that $q$ is provable defeasibly
we have two choices: (1) We show that $q$ is already definitely
provable; or (2) we need to argue using the defeasible part of $\T$ as
well. In particular, we require that there must be a strict or
defeasible rule with head $q$ which can be applied (2.1). But now we
need to consider possible ``attacks'', that is, reasoning chains
in support of $\non q$. To be more specific: to prove $q$ defeasibly
we must show that $\non q$ is not definitely provable
(2.2). Also (2.3) we must consider the set of all  rules which are not
known to be inapplicable and which have 
head $\non q$ (note that here we consider defeaters, too, whereas
they could not be used to support the conclusion $q$; this is in line
with the motivation of defeaters given earlier).
Essentially each such rule $s$ attacks the conclusion $q$. For
$q$ to be provable, each such rule $s$ must be counterattacked by a
rule $t$ with head $q$ with the following properties: (i) $t$ must be
applicable at this point, and (ii) $t$ must be stronger than $s$. Thus
each attack on the conclusion $q$ must be counterattacked by a
stronger rule. 

From $T_\tit{bird}$ in Example \ref{bird}
we can infer $\pd{\tit{bird}(\tit{ethel})}$ in a proof of length 3, using
part (1) of the $\pd{}$ inference rule.
we can infer $\pd{\tit{bird}(\tit{ethel})}$ in a proof of length 3, using (1).
Other applications of this inference rule first require
application of the inference rule $-\partial$.
This inference rule completes
the definition of the proof system of defeasible logic.
It is a strong negation of the inference rule $+\partial$ \cite{ABGM}.

\begin{tabbing}
90123456\=7890\=1234\=5678\=9012\=3456\=\kill

\>$-\partial$: \> We may append $P(i+1)=-\partial q$ if  \\
\>\>(1) $-\Delta q \in P(1..i)$ and \\
\>\>(2) \> (2.1) $\forall r\in R_{\tit{sd}}[q] \ \exists a \in
A(r): -\partial a\in P(1..i)$ or \\
\>\>\>(2.2) $+\Delta\non q \in P(1..i)$ or \\
\>\>\>(2.3) $\exists s \in R[\non q]$ such that \\
\>\>\>\>(2.3.1) $\forall a\in A(s): +\partial a\in P(1..i)$ and \\
\>\>\>\>(2.3.2) $\forall t\in R_{\tit{sd}}[q]$ either \\
\>\>\>\>\> $\exists a\in A(t): -\partial a\in P(1..i)$ or $t\not> s$
\end{tabbing}

To prove that $q$ is not defeasibly provable, we must first establish
that it is not definitely provable. Then we must establish that it
cannot be proven using the defeasible part of the theory. There are
three possibilities to achieve this: either we have established that
none of the (strict and defeasible) rules with head $q$ can be applied
(2.1); or $\non q$ is definitely provable (2.2); or there must be an
applicable rule $s$ with head $\non q$ such that no possibly applicable rule
$t$ with head $q$ is superior to $s$ (2.3).

From $T_\tit{bird}$ in Example \ref{bird}
we can infer $\md{\tit{brokenWing}(\tit{ethel})}$ with a proof of length 2
using (1) and (2.1),
since $R[q]$ is empty.
Employing this conclusion, we can
then infer $\md{\neg \tit{flies}(\tit{ethel})}$, again using (2.1),
since the only rule $r$ is $r_{4,e}$.
Using $r_5$,
we can infer $\pd{\tit{heavy}(\tit{ethel})}$ by (2) of the inference rule $\pd{}$;
(2.3) holds since $R[\non q]$ is empty.
Using (2.3) of the inference rule $\md{}$,
we can infer $\md{\tit{flies}(\tit{ethel})}$,
where the role of $s$ is taken by $r_{3,e}$.

Furthermore,
we can infer $\md{\tit{brokenWing}(\tit{tweety})}$ and $\md{\tit{heavy}(\tit{tweety})}$
using (2.1) of inference rule $\md{}$,
since in both cases $R_{\tit{sd}}[q]$ is empty.
Employing those conclusions,
we can then infer $\md{\neg \tit{flies}(\tit{tweety})}$, again using (2.1),
since the only rules for $\neg \tit{flies}(\tit{tweety})$ involve antecedents
already established to be unprovable defeasibly.
Now, using this conclusion, we can use (2), specifically (2.1), of
the inference rule $\pd{}$ to infer $\pd{\tit{flies}(\tit{tweety})}$.

We write $\T \vdash c$,
where $\T$ is a defeasible theory and $c$ is a conclusion,
if there is a proof in $\T$ that ends with $c$.
For example, $\T_\tit{bird} \vdash \pd{\tit{flies}(\tit{tweety})}$.

It is clear that \tit{DL} is a paraconsistent logic.
If a defeasible theory $\T$ contains facts $a$ and $\neg a$ and
a rule $b \rightarrow b$, then the inconsistency concerning $a$
has no effect on what is provable about $b$.
When considering defeasible reasoning,
a main feature of defeasible logics is that
potentially inconsistent inferences are avoided either
by use of the superiority relation or by simply failing to infer
either of the consequences, as in the previous discussion
of a bird with a broken wing.
Some inconsistent defeasible inferences may be made,
as consequences of inconsistent definite inferences,
but the logic displays the same kind of paraconsistency
as discussed above for definite reasoning.

\section{A Model-Theoretic Semantics}

A domain is a $\Sigma$-structure $\D$ that
defines a set $D$ of objects over which formulas may be interpreted,
and gives a meaning to all function symbols and
pre-defined predicate symbols
as functions (respectively, relations) over $D$.
$=$ is interpreted as identity on $D$.
Given a domain $\D$, the $\D$-base of a language is
$B_\D = \{p(d_1, \ldots, d_n), \neg p(d_1, \ldots, d_n) ~|~ 
  n \geq 0, p \in \Pi, p \mbox{ has arity } n,
  d_i \in D \mbox{ for } i = 1,2,\ldots,n \}$.
We sometimes refer to elements of $B_\D$ as interpreted literals.

The following definition introduces the notion of a defeasible interpretation.
It is formulated in a truth-functional manner to simplify later statements,
but it might equally well be formulated without introducing truth,
at the expense of clumsier definitions later.
\begin{definition}
A {\em defeasible interpretation} $\I$ consists of a domain $\D_\I$,
and two partial functions $\I_\Delta$ and $\I_\partial$
which map $B_{\D_\I}$ to $\{\tit{True}, \tit{False}\}$,
which satisfy the following conditions\footnote{
Actually these conditions need not be imposed:
even if we did not impose them on interpretations,
all models of a defeasible theory satisfy these conditions.
But we prefer to explicitly require that interpretations respect
this fundamental relationship between definite and defeasible knowledge.
}:

\[
\forall q \in B_{\D_\I} ~ ~
\I_\Delta(q) = \tit{True} \rightarrow \I_\partial(q) = \tit{True} 
\]
and
\[
\forall q \in B_{\D_\I} ~ ~
\I_\partial(q) = \tit{False} \rightarrow \I_\Delta(q) = \tit{False} 
\]

A $\D$-interpretation is a defeasible interpretation over the domain $\D$.
\end{definition}

\newcommand{\ID}{{\I_\Delta}}
\newcommand{\Id}{{\I_\partial}}
\newcommand{\MDel}{{\M_\Delta}}
\newcommand{\Md}{{\M_\partial}}

Intuitively, 
$\I_\Delta$ specifies those evaluated literals that are known
definitely (map to $\tit{True}$)
or known not to be known definitely (map to $\tit{False}$).
Similarly,
$\I_\partial$ specifies those evaluated literals that are believed
defeasibly (map to $\tit{True}$)
or known not to be defeasibly believed (map to $\tit{False}$).
Thus a defeasible interpretation represents a state of mind in
which some evaluated literals are known or believed,
while for others there is awareness that they are not known or believed,
and still others have undefined status --
neither known, nor unknown;
neither believed, nor unbelieved.
We refer to the values of $\I_\Delta$ and $\I_\partial$ 
on an evaluated literal $q$ as the {\em epistemic status} of $q$ in $\I$.

The two conditions on defeasible interpretations enforce 
that that which is known is believed, and
that which is not believed is not known.
In other words, they enforce the expected relationship
between knowledge and belief.

The functions $\ID$ and $\Id$ are easily extended to conjunctions
of evaluated literals.
The extension follows the same pattern for both functions, so
we show only $\ID$.
\[
\ID(q_1,\ldots,q_n) = \bigwedge_{i=1}^n \ID(q_i)
\]
where the truth table for $\wedge$ is

\begin{center}
\begin{tabular}{c|c|c|c}
              &   $\tit{True}$ & $~~~~\tit{False}~~~~$ & ~~undefined~~   \\ \hline
     $\tit{True}$   & $\tit{True}$   & $\tit{False}$         & ~~undefined~~   \\ 
     $\tit{False}$  & $\tit{False}$  & $\tit{False}$         & $\tit{False}$ \\ 
     ~~undefined~~  & ~~undefined~~ & $\tit{False}$ & ~~undefined~~   \\ 
\end{tabular}
\end{center}
\vspace{0.5cm}

\finish{

comments on Kripke/Kleene etc, check ofp paper, Fitting paper
}

Among the $\D$-interpretations,
we distinguish the {\em $\D$-models} of a defeasible theory $\T$:
those $\D$-interpretations which are consistent with
the rules of reasoning specified by $\T$ and \tit{DL}.

\begin{definition}
A $\D$-interpretation $\I$ is a {\em $\D$-model} if
it satisfies the following four conditions.
Let $q$ range over elements of $B_\D$.

\begin{tabbing}
90123456\=7890\=1234\=5678\=9012\=3456\=\kill

$\ID(q) = \tit{True}$ iff \\
\>\>\> $q\in F$ or \\
\>\>\> $\exists r\in R_{s}[q] \ \ID(A(r)) = \tit{True}$
\end{tabbing}

\begin{tabbing}
90123456\=7890\=1234\=5678\=9012\=3456\=\kill

$\ID(q) = \tit{False}$ iff \\
\>\>\> $q\not\in F$ and \\
\>\>\> $\forall r\in R_{s}[q] \ \ID(A(r)) = \tit{False}$
\end{tabbing}

\begin{tabbing}
90123456\=7890\=1234\=5678\=9012\=3456\=\kill
$\Id(q) = \tit{True}$ iff \\

\>\>(1) $\ID(q) = \tit{True}$ or \\
\>\>(2) \> (2.1) $\exists r\in R_{\tit{sd}}[q] \ \Id(A(r)) = \tit{True}$ and \\
\>\>\>(2.2) $\ID(\non q) = \tit{False}$ and \\
\>\>\>(2.3) $\forall s \in R[\non q]$ either \\
\>\>\>\>(2.3.1) $\Id(A(s)) = \tit{False}$ or \\
\>\>\>\>(2.3.2) $\exists t\in R_{\tit{sd}}[q]$ such that \\
\>\>\>\>\> $\Id(A(t)) = \tit{True}$ and $t>s$ 
\end{tabbing}

\begin{tabbing}
90123456\=7890\=1234\=5678\=9012\=3456\=\kill
$\Id(q) = \tit{False}$ iff \\

\>\>(1) $\ID(q) = \tit{False}$ and \\
\>\>(2) \> (2.1) $\forall r\in R_{\tit{sd}}[q] \ \Id(A(r)) = \tit{False}$ or \\
\>\>\>(2.2) $\ID(\non q) = \tit{True}$ or \\
\>\>\>(2.3) $\exists s \in R[\non q]$ such that \\
\>\>\>\>(2.3.1) $\Id(A(s)) = \tit{True}$ and \\
\>\>\>\>(2.3.2) $\forall t\in R_{\tit{sd}}[q]$ either \\
\>\>\>\>\> $\Id(A(t)) = \tit{False}$ or $t\not>s$ 
\end{tabbing}
\end{definition}

Clearly there is a close correspondence between these conditions
and the inference rules of \tit{DL}.
The four conditions are if-and-only-if statements.
The ``if'' part ensures that all models are deductively closed.
The ``only-if'' part ensures that models are abductively closed,
in the following sense:
if an evaluated literal has some epistemic status in a model,
then there is a reason, in the model, why that status is given.

We must also represent the meaning of the symbols of $\Sigma$.
We might simply use a $\Sigma$-structure $\D$ for this purpose,
but it turns out that, in general,
models based on the intended $\Sigma$-structure $\D$ do not characterize
provability in \tit{DL}.

Instead we will use the first-order theory of $\D$, denoted $Th(\D)$.
Thus we will consider all domains that agree with $\D$
on first-order sentences.
We write $Th(\D), \T \models_{\tit{DL}} c$ to denote that
the conclusion $c$ holds in all defeasible models $M$ of $\T$
such that $\D_M$ is a model of $Th(\D)$
(in the conventional sense of first-order logic).
That is, if $c$ is $\PD{q}$ then $M_\Delta(q) = \tit{True}$,
         if $c$ is $\MD{q}$ then $M_\Delta(q) = \tit{False}$, etc.
We say that $c$ is a {\em defeasible logical consequence} of $\T$ and $Th(\D)$.

Taking the defeasible models of $Th(\D), \T$ as the semantics of 
the defeasible theory $\T$ over $\D$,
it is straightforward to prove that the proof system is sound.

\begin{theorem}
Let $\T$ be a defeasible theory over a domain $\D$.
Let $c$ be a conclusion.

If $\T \vdash c$ then
$Th(\D), \T \models_{\tit{DL}} c$
\end{theorem}
\ignore{
\begin{proof}

The proof is by induction on the length of proofs.
We present only the details in the case where $c$ is $\pd{q}$
for some literal $q$.
The other cases follow a similar pattern.

Suppose $c$ is $\pd{q}$ and $\T \vdash c$
with a proof $P$ of length $k+1$.
Then, from the inference rule $\pd{}$,
either $\PD{q} \in P(1..k)$; or
$\MD{\non q} \in P(1..k)$ and
there is a rule $r \in R_{\tit{sd}}[q]$ such that 
$\forall a \in A(r): +\partial a\in P(1..k)$ 
and for every rule $s \in R[\non q]$ either
$\exists a\in A(s): -\partial a\in P(1..k)$ 
or there is a rule $t \in R_{\tit{sd}}[q]$ such that 
$\forall a\in A(t): +\partial a\in P(1..k)$ 
and $t>s$.

Hence, by the induction hypothesis,
in every model $\M$ of $Th(\D), \T$,
either $\MDel(q) = \tit{True}$ or
$\MDel(\non q) = \tit{False}$ and
there is a rule $r \in R_{\tit{sd}}[q]$ such that 
$\Md{A(r)} = \tit{True}$
and for every rule $s \in R[\non q]$ either
$\Md{A(s)} = \tit{False}$
or there is a rule $t \in R_{\tit{sd}}[q]$ such that 
$\Md{A(t)} = \tit{True}$
and $t>s$.

But then, by the third condition on models 
$\Md(q) = \tit{True}$.

\end{proof}
}

The proof is by induction on the length of proofs.

\section{Completeness}  \label{sec:complete}

We now establish the  completeness of the
inference rules with respect to these models.
We do this by establishing an intimate connection between
$\D$-models of a defeasible theory $\T$ and
partial models of a corresponding metaprogram.
In this section we only outline the proof,
since there is not enough space to present the requisite background material.

In \cite{MG99} we developed a logic metaprogram $\M$ for \tit{DL}.
If $\T$ is a defeasible theory,
then $\MT$ is $\M$ augmented with a representation of $\T$.
It was shown in \cite{MG99} that
the consequences of $\T$ are characterized by $\MT$
under Kunen's semantics for logic programs \cite{Kunen}.

\begin{theorem}[\cite{MG99}]  \label{thm:proofs}
Let $\T$ be a defeasible theory
and let \MT denote its metaprogram counterpart.
Let $\H$ be the Herbrand domain over which $\MT$ is defined.

  For each literal $p$,
  \begin{enumerate}
  \item $\T\vdash+\Delta p$ iff
    $\MT, Th(\H) \vdash_{K}\mt{definitely}(p)$;
  \item $\T\vdash-\Delta p$ iff
    $\MT, Th(\H) \vdash_{K}\mt{\neg definitely}(p)$;
  \item $\T\vdash+\partial p$ iff
    $\MT, Th(\H) \vdash_{K}\mt{defeasibly}(p)$;
  \item $\T\vdash-\partial p$ iff
    $\MT, Th(\H) \vdash_{K}\mt{\neg defeasibly}(p)$;
  \end{enumerate}
\end{theorem}

Kunen also characterized his semantics of logic programs
in terms of 3-valued models of the logic program.
For any logic program $\P$ and literal $q$,
Kunen showed that
$\P, Th(\H) \vdash_K q$ iff $\P, Th(\H) \models_K q$ \cite{Kunen},
where $\models_{K}$ denotes the 3-valued logical consequence relation.

The remaining step is to relate defeasible models of a defeasible theory $\T$
with 3-valued models of $\MT$.
There is a technical difficulty:
defeasible interpretations of $\T$ involve domains with signature $\Sigma$,
whereas the signature of domains of $\MT$
extends $\Sigma$ with function symbols corresponding to $\Pi$
(and other, auxiliary function symbols).
The problem arises because the metaprogram represents
predicates as functions.

We address this problem by defining,
for every domain $\D$ over $\Sigma$, a domain extension $\D^*$
over the extended signature.
We are then able to relate
$\D$-interpretations of $\T$ with
3-valued interpretations of $\MT$ over $\D^*$.
Thus we obtain

\begin{proposition}  \label{prop:logcons}
Let $\T$ be a defeasible theory over signature $\Sigma$
and let $\D$ be a domain over this signature.

\begin{itemize}
\item
$\MT, Th(\D^*) \models_K \mt{definitely}(q)$ iff $\T, Th(\D) \models_{\tit{DL}} \PD{q}$
\item
$\MT, Th(\D^*) \models_K \neg \mt{definitely}(q)$ iff $\T, Th(\D) \models_{\tit{DL}} \MD{q}
$
\item
$\MT, Th(\D^*) \models_K \mt{defeasibly}(q)$ iff $\T, Th(\D) \models_{\tit{DL}} \pd{q}$
\item
$\MT, Th(\D^*) \models_K \neg \mt{defeasibly}(q)$ iff $\T, Th(\D) \models_{\tit{DL}} \md{q}
$
\end{itemize}
\end{proposition}

Combining these three results together,
the completeness (and soundness)
of the proof system for the model theory is established.

\begin{theorem}
Let $\T$ be a defeasible theory over signature $\Sigma$
and let $\H$ be the Herbrand domain over this signature.
Let $c$ be any conclusion.

$\T \vdash c$ iff $Th(\H), \T \models_{\tit{DL}} c$
\end{theorem}

\finish{
\subsection{The Defeasible Logic Metaprogram}

In this section we outline a metaprogram \M in a logic programming form
that expresses the essence of the defeasible reasoning
embedded in the proof theory.
It was introduced in \cite{MG99}.
This metaprogram is a key intermediate step in establishing
the completeness of the proof theory.
The metaprogram assumes that the following predicates,
which are used to represent a defeasible theory, are defined.
\begin{itemize}
\item $\mt{fact}(Head)$,

\item $\mt{strict}(Label,Head,Body)$,

\item $\mt{defeasible}(Label,Head,Body)$,

\item $\mt{defeater}(Label,Head,Body)$, and 
\item $\mt{sup}(Rule1,Rule2)$, 
\end{itemize}
\M consists of the following clauses.
We first introduce the predicates defining classes of rules, namely
\begin{clause}
$\mt{supportive\_rule}(Label,Head,Body)$:-\\
  \> $\mt{strict}(Label,Head,Body)$.\\
\\[-.5\baselineskip]
$\mt{supportive\_rule}(Label,Head,Body)$:-\\
  \> $\mt{defeasible}(Label,Head,Body)$.
\end{clause}

\begin{clause}
  $\mt{rule}(Label,Head,Body)$:-\\
  \> $\mt{supportive\_rule}(Label,Head,Body)$.\\
\\[-.5\baselineskip]
  $\mt{rule}(Label,Head,Body)$:-\\
  \> $\mt{defeater}(Label,Head,Body)$.  
\end{clause}
We introduce now the clauses defining the predicates corresponding to
$\PD{}$, $\MD{}$, $\pd{}$, and $\md{}$.
These clauses specify the structure of defeasible reasoning in Defeasible Logic.
Arguably they convey the conceptual simplicity of Defeasible Logic
more clearly than does the proof theory.
\begin{Clause}\label{strictly1}
  $\mt{definitely}(X)$:-\\
  \> $\mt{fact}(X)$.
\end{Clause}

\begin{Clause}\label{strictly2}
  $\mt{definitely}(X)$:-\\
  \> $\mt{strict}(R,X,[\seq{Y}])$,\\
  \> $\mt{definitely}(Y_1)$, \dots ,$\mt{definitely}(Y_n)$.
\end{Clause}

\begin{Clause}\label{defeasibly1}
  $\mt{defeasibly}(X)$:-\\
  \> $\mt{definitely}(X)$.
\end{Clause}

\begin{Clause}\label{defeasibly2}
  $\mt{defeasibly}(X)$:-\\
  \> $\mt{not\ definitely}(\sim X)$,\\
  \> $\mt{supportive\_rule}(R,X,[\seq{Y}])$,\\
  \> $\mt{defeasibly}(Y_1)$, \dots ,$\mt{defeasibly}(Y_n)$,\\
  \> $\mt{not\ overruled}(R,X)$.
\end{Clause}

\begin{Clause}\label{overruled}
  $\mt{overruled}(R,X)$:-\\
  \> $\mt{rule}(S,\sim X,[\seq{U}])$,\\
  \> $\mt{defeasibly}(U_1)$, \dots ,$\mt{defeasibly}(U_n)$,\\
  \> $\mt{not\ defeated}(S,\sim X)$.
\end{Clause}

\begin{Clause}\label{defeated}
  $\mt{defeated}(S,\sim X)$:-\\
  \> $\mt{sup}(T,S)$, \\
  \> $\mt{supportive\_rule}(T,X,[\seq{V}])$,\\
  \> $\mt{defeasibly}(V_1)$, \dots ,$\mt{defeasibly}(V_n)$.
\end{Clause}

The first two clauses address definite provability, while the
remainder address defeasible provability. 
The clauses specify if and how a rule in Defeasible Logic can be over-ridden
by another, and which rules can be used to defeat an over-riding rule,
among other aspects of the structure of defeasible reasoning.

We have permitted ourselves some syntactic flexibility
in presenting the metaprogram.
However, there is no technical difficulty in using conventional
logic programming syntax to represent this program.

Given a defeasible theory $\T=(F,R,>)$, the
corresponding program \MT is obtained from \M by adding facts
according to the following guidelines:
\begin{enumerate}
\item $\mt{fact}(p)$.  \hfill for each $p\in F$;
\item $\mt{strict}(r_i,p,[q_1,\dots,q_n])$. \\
\hspace*{\fill}
for each rule
  $r_i:q_1,\dots,q_n\to p\in R$;
\item $\mt{defeasible}(r_i,p,[q_1,\dots,q_n])$. \\
\hspace*{\fill}
for each rule
  $r_i:q_1,\dots,q_n\Rightarrow p\in R$;
\item $\mt{defeater}(r_i,p,[q_1,\dots,q_n])$. \\
\hspace*{\fill}
for each rule 
  $r_i:q_1,\dots,q_n\leadsto p\in R$;
\item $\mt{sup}(r_i,r_j)$. \\
\hspace*{\fill}
for each pair of rules such that
  $r_i > r_j$.
\end{enumerate}

\subsection{Kunen Semantics}
Kunen's semantics \cite{Kunen}
is a 3-valued semantics for logic programs.
It was introduced as a formalization of a notion of negation-as-failure
that is a practical idealization of
negation-as-failure as implemented in Prolog.

A {\em 3-valued interpretation}
is a mapping from interpreted atoms to one of three truth values:
$\mathbf{t}$ (true), $\mathbf{f}$ (false), and $\mathbf{u}$ (unknown).
This mapping can be extended to all formulas using Kleene's
3-valued logic.

Kleene's truth tables can be summarized as follows. If $\phi$ is a
boolean combination of the truth values $\mathbf{t}$, $\mathbf{f}$, and
$\mathbf{u}$, its truth value is $\mathbf{t}$ iff all the possible
ways of putting in $\mathbf{t}$ or $\mathbf{f}$ for the various
occurrences of $\mathbf{u}$ lead to a value $\mathbf{t}$ being computed
in ordinary 2-valued logic: $\phi$ gets the value $\mathbf{f}$ iff
$\neg\phi$ gets the value $\mathbf{t}$ by this method, 
and $\phi$ gets the value
$\mathbf{u}$ otherwise. These truth values can be extended in the
obvious way to predicate logic, thinking of the quantifiers as
infinite disjunction or conjunction.
A 3-valued interpretation $I$ is a 3-valued model for a sentence $\phi$
if $I(\phi) = \mathbf{t}$.

The Kunen semantics of a program $\P$ is obtained from a sequence $\{ I_n \}$
of 3-valued interpretations, defined as follows.

\begin{enumerate}
\item $I_0(\alpha) = \mathbf{u}$ for every interpreted atom $\alpha$
\item $I_{n+1}(\alpha)=\mathbf{t}$ iff for some clause
  $$\beta\hbox{:-}\phi$$
  in the program, $\alpha=\beta\sigma$ for some valuation
  $\sigma$ for variables in the clause such that
  $$I_n(\phi\sigma)=\mathbf{t}\ .$$
\item $I_{n+1}(\alpha)=\mathbf{f}$ iff for all the clauses
  $$\beta\hbox{:-}\phi$$
  in the program, and all valuations $\sigma$, if
  $\alpha=\beta\sigma$, then
  $$I_n(\phi\sigma)=\mathbf{f}\ .$$
\item $I_{n+1}(\alpha)=\mathbf{u}$ otherwise.
\end{enumerate}
We shall say that the Kunen semantics of $\P$ supports $\alpha$
iff there is an interpretation $I_n$, for some finite $n$,
such that $I_n(\alpha)=\mathbf{t}$.
We use $\P\vdash_K\alpha$ to denote that the Kunen semantics for
the program $\P$ supports $\alpha$.

Our reason for introducing Kunen's semantics is the following result,
which shows the equivalence of inference in \tit{DL} and
inference by Kunen's semantics from the metaprogram.
In particular, it shows that the notion of failure-to-prove
that is present in the inference rules $\MD{}$ and $\md{}$
is characterized by Kunen's semantics of negation-as-failure.

\begin{theorem}[\cite{MG99}]  \label{thm:proofs}
Let $\T$ be a defeasible theory
and let \MT denote its metaprogram counterpart.
Let $\H$ be the Herbrand domain over which $\MT$ is defined.

  For each literal $p$,
  \begin{enumerate}
  \item $\T\vdash+\Delta p$ iff 
    $\MT, Th(\H) \vdash_{K}\mt{definitely}(p)$;
  \item $\T\vdash-\Delta p$ iff 
    $\MT, Th(\H) \vdash_{K}\mt{\neg definitely}(p)$;
  \item $\T\vdash+\partial p$ iff 
    $\MT, Th(\H) \vdash_{K}\mt{defeasibly}(p)$;
  \item $\T\vdash-\partial p$ iff 
    $\MT, Th(\H) \vdash_{K}\mt{\neg defeasibly}(p)$;
  \end{enumerate}
\end{theorem}

Thus Kunen's semantics of $\MT$ characterizes the consequences of
$\T$ in \tit{DL}.

Kunen provided an equivalent characterization of his semantics
in terms of 3-valued models of the Clark completion \cite{Clark}
of a logic program.
For a logic program $\P$,
we write $\P \models_K q$ if $q$ is true in every 3-valued model
of the Clark-completion of $\P$.
For any logic program $\P$ and literal $q$,
Kunen showed that
$\P, Th(\H) \vdash_K q$ iff $\P, Th(\H) \models_K q$ \cite{Kunen}.

\subsection{Correspondence}
We will now establish a correspondence between
defeasible models of a defeasible theory $\T$ and
3-valued models of $\MT$.
This correspondence will lead us directly to 
the completeness result we desire.

First we need to define the domain extension $\D^*$ of a domain $\D$.
$\D^*$ is a $(\Sigma \cup \Pi')$-structure,
where $\Sigma$ is the signature of $\D$ and
$\Pi'$ includes an $n$-ary function symbol for each $n$-ary predicate symbol
in $\Pi$.  
$\Pi'$ also contains a list constructor and an empty list,
and a unary symbol representing negation in \tit{DL}.
The carrier is
$D^* = \{ t(\d) ~|~ t(\x) \mbox{ is a $\Pi'$-term over $n$ variables and } \d \in D^n \}$.
For an $n$-ary function symbol $f \in \Pi'$,
$f(\d)$ is the term constructed with label ``$f$'' and children $\d$.
For an $n$-ary function symbol $f \in \Sigma$,
$f(\d) = d'$ if $\d \in D^n$ and $f(\d) = d'$ in $\D$;
if $\d \not\in D^n$ then 
$f(\d)$ is the term constructed with label ``$f$'' and children $\d$.
$d_1 = d_2$ iff $d_1$ and $d_2$ are identical.
For any other $n$-ary predicate symbol $p \in \Sigma$,
$p(\d)$ is true iff $\d \in D^n$ and $p(\d)$ is true in $\D$.

We can now define correspondence between
defeasible interpretations of $\T$ and 3-valued interpretations of $\MT$.

\begin{definition}
Consider interpretations with domain $\D$.

Given a defeasible interpretation $N$ of $\T$ over $\D$,
a 3-valued interpretation $N'$ of $\MT$ over $\D^*$
{\em corresponds} to $N$ if the following statements hold:
\begin{itemize}
\item
$\forall q \in B_\D ~ N_\Delta(q) = \tit{True} ~\iff~ N'(definitely(q)) = \mathbf{t}$
\item
$\forall q \in B_\D ~ N_\Delta(q) = \tit{False} ~\iff~ N'(definitely(q)) = \mathbf{f}$
\item
$\forall q \in B_\D ~ N_\partial(q) = \tit{True} ~\iff~ N'(defeasibly(q)) = \mathbf{t}$
\item
$\forall q \in B_\D ~ N_\partial(q) = \tit{False} ~\iff~ N'(defeasibly(q)) = \mathbf{f}$
\end{itemize}
Similarly,
given a 3-valued interpretation $N'$ of $\MT$ over $\D^*$,
a defeasible interpretation $N$ of $\T$ over $\D$
{\em corresponds} to $N'$ if the previous statements hold.
\end{definition}
Notice that every 3-valued interpretation has a unique corresponding
defeasible interpretation.

We can now establish that the defeasible models of $T$
and the 3-valued models of $\MT$ correspond.

\begin{lemma}  \label{lem:corresp}
Let $\T$ be a defeasible theory over signature $\Sigma$
and let $\D$ be a domain over this signature.
Then
\begin{itemize}
\item
For any defeasible model of $\T$ over $\D$
there is a corresponding 3-valued model of $\MT$ over $\D^*$.
\item
For any 3-valued model of $\MT$ over $\D^*$,
there is a corresponding defeasible model of $\T$ over $\D$
\end{itemize}
\end{lemma}
With the help of this lemma, we can establish that
defeasible logical consequences of $\T$
correspond to 3-valued logical consequences of $\MT$.

\begin{proposition}  \label{prop:logcons}
Let $\T$ be a defeasible theory over signature $\Sigma$
and let $\D$ be a domain over this signature.

\begin{itemize}
\item
$\MT, Th(\D^*) \models_K definitely(q)$ iff $\T, Th(\D) \models_{\tit{DL}} \PD{q}$
\item
$\MT, Th(\D^*) \models_K \neg definitely(q)$ iff $\T, Th(\D) \models_{\tit{DL}} \MD{q}$
\item
$\MT, Th(\D^*) \models_K defeasibly(q)$ iff $\T, Th(\D) \models_{\tit{DL}} \pd{q}$
\item
$\MT, Th(\D^*) \models_K \neg defeasibly(q)$ iff $\T, Th(\D) \models_{\tit{DL}} \md{q}$
\end{itemize}
\end{proposition}
Using this and earlier results,
we obtain the soundness and completeness of the 
inference rules of \tit{DL} for obtaining defeasible logical consequences.
The proof is a combination of
Theorem \ref{thm:proofs},
Kunen's dual characterization of his semantics of logic programs,
and Proposition \ref{prop:logcons}
\begin{theorem}
Let $\T$ be a defeasible theory over signature $\Sigma$
and let $\H$ be the Herbrand domain over this signature.
Let $c$ be any conclusion.

$\T \vdash c$ iff $Th(\H), \T \models_{\tit{DL}} c$
\end{theorem}
}

\section{Other Defeasible Logics}

The approach of this paper is more widely applicable than simply to \tit{DL}.
In the following, we first look at the extension of this approach
to a form of \tit{DL} that is not essentially propositional.
We then discuss its application to other defeasible logics that
can be formulated within the flexible framework of \cite{ABGM}.
We only have space to outline this work.

\subsection{Non-Propositional Defeasible Logic}

The development of the model theory in terms of $\D$ and $Th(\D)$
is an overkill, technically,
for an essentially propositional logic,
such as the presentation of \tit{DL} in Section \ref{sec:DL}.
However, this development makes it easy to extend \tit{DL}
to a first-order logic, and to extend it further
with pre-defined functions and predicates --
constraints in the sense of constraint logic programming \cite{JM}.

The only change to the model theory is that the restriction
on $\Sigma$ (to constants and =) is removed,
with consequent flow-on affects on $\D$.

The proof system, as presented in Section \ref{sec:DL},
is inadequate to handle infinite domains,
if rules are regarded as schemas for propositional rules.
However \cite{MG99} gives a bottom-up reformulation of the proof system
that can handle infinite domains, and is easily extended
to address constraints.

With this reformulated proof system,
the soundness and completeness continues to hold
in the presence of constraints over infinite domains.
There are some elements of the proofs that must change.
In particular,
the metaprogram $\M$ must now accommodate constraints
and Theorem \ref{thm:proofs}
must be extended.
But such changes are not difficult.
The extension of Kunen's result, equating $\vdash_K$ and $\models_K$,
to arbitrary domains is due to Stuckey \cite{Stuckey}.

\subsection{Variants of \tit{DL}}

As proposed in \cite{ABGM},
the metaprogram presentation of \tit{DL} can be generalized to a framework
for the definition of many different defeasible logics.
\cite{ABGM} presents some variants of \tit{DL} 
with different proof-theoretic properties.
Some of these variants involve extra levels of belief,
beyond definite and defeasible knowledge.

Since failure-to-prove in these variants is still characterized
by Kunen's semantics, we can directly apply the techniques
of this paper to establish a model theory and prove
soundness and completeness of these variants.
Technically, all that is needed is
a definition of model appropriate to the inference rules of the logic
and a corresponding metaprogram.

The framework of \cite{ABGM} admits other forms of failure-to-prove
than Kunen's semantics.  Any semantics for logic programs provides
an alternate form of failure-to-prove.
Indeed, \cite{MG99} showed that Well-Founded Defeasible Logic
is characterized by metaprogram of Section \ref{sec:complete} and
the well-founded semantics of logic programs \cite{VGRS}.

In the well-founded semantics 
-- and many other semantics of logic programs --
the semantics rests on a single domain of values $\D$.
Thus, in these cases, the introduction of $Th(\D)$ is unnecessary.
In the case of Well-Founded Defeasible Logic there is,
essentially, a single model of a defeasible theory $\T$ over $\D$,
which is the reduct of the well-founded (partial) model of $\MT$.
The defeasible logic investigated in \cite{Donnelly,ND}
performs loop-checking to detect failure-to-prove,
which makes failure-to-prove in that logic similar to
well-founded semantics.
This suggests that a model-theoretic semantics for which that logic
is sound and complete might be established by following the pattern
of this paper, but using the well-founded semantics in place of
Kunen's semantics and using a substantially different metaprogram.

Other semantics, such as the stable model semantics \cite{GL88},
give rise to several $\D$-models for a defeasible theory
in the corresponding defeasible logic.

\section{Conclusion}

We have introduced an approach to defining an appropriate model theory
for defeasible logics.  The approach was demonstrated in detail
for the logic \tit{DL}, and we outlined how it can be applied to
a wide range of defeasible logics.

\section*{Acknowledgements}

Thanks to
D. Billington,
who introduced me to the problem of defining a model theory for defeasible logic,
and gave me access to his extensive library of papers on defeasible logic,
and to
G. Antoniou and G. Governatori
for discussions on defeasible logic.
This research was supported by  the
Australian Research Council.

\end{document}